\documentclass{nature2}
\usepackage[english]{babel}
 
\usepackage[numbers,super,compress]{natbib}
\usepackage[pdftex]{graphicx}
\usepackage{caption}
\usepackage{amsmath,amssymb}
\usepackage[]{pdfpages}
\usepackage{enumerate}
\usepackage[normalem]{ulem}
\usepackage{hyperref}

\def\deg{\ifmmode{^\circ}\else{$^\circ$}\fi}
\def\co2{\ifmmode{{\rm CO}_2}\else{CO$_2$}\fi}
\def\h2o{\ifmmode{{\rm H}_2{\rm O}}\else{H$_2$O}\fi}

\begin{document}

\title{The Martian mid-latitude subsurface ice is the remnant of a past ice sheet}
\date{}




\author{
\renewcommand{\thefootnote}{\arabic{footnote}}
 E. Vos
\hspace{-.5ex}\footnote{LMD, Institut Pierre Simon Laplace Universit\'{e} Paris 6, France.
\newline \hspace*{0.25in}Email: \href{mailto:Eran.Vos@gmail.com}{Eran.Vos@gmail.com}}$^1$, F. Forget$^{1}$, L. Lange$^1$ ,
   J. Naar$^1$, J.B. Clement$^1$, E. Millour$^1$}

\maketitle

\begin{abstract}
On Mars, a relatively pure water ice layer lies beneath several centimeters of dry soil at midlatitudes. Its widespread presence poleward of 60\deg{} latitude was detected by remote neutron spectroscopy\citep{Boynton2002} and confirmed by the Phoenix lander \citep{Mellon2009} at 68\deg{}N. Recent observations of exposed ice \citep{dundas2023,Dundas2021,Byrne2009b} indicate that the near-surface ice layer extends to 35\deg{} latitude and exhibits pronounced spatial structure. However, previous models \citep{MELLON2022,Schorghofer2012,Bramson2017,lange2023} did not capture the observed spatial structure of the midlatitude ice layer. 
Here, on the basis of improved calculations using the Mars Planetary Climate Model \citep{forget1999}, we show that mid-latitude buried ice could be the remnant of a ice layer deposited on the surface when the obliquity was higher than today\citep{madeleine2014,naar2024recent}. Assuming that the ice subsequently sublimated and became buried beneath a sublimation lag, we estimate that surface ice emplaced 630 kyr (~4.18 Myr) ago at 35\textdegree~obliquity (40\textdegree~), at latitudes of 40–55\textdegree N, would today reside at depths of 25–150 (41–255) cm, depending on the regolith and ice properties. For ice emplaced 630 kyr ago, the modeled burial depths align with observations and capture the observed longitudinal depth variations  \citep{Morgan2021}, in contrast to ice emplaced 4.18 Myr ago. We therefore infer that the mid-latitude subsurface ice is younger than 4 Myr.

\end{abstract}

\section*{}

On Mars, Subsurface water ice can be emplaced in two ways. It may form as a surface layer, for example through direct atmospheric deposition or snowfall, which then sublimates and buries itself beneath a dust lag left behind by the small amount of impurities originally contained in the ice. Alternatively, ice may accumulate directly within subsurface pores by vapor diffusion \citep{mellon1993,hudson2009,schorghofer2005}. Nearly pure ice is much more consistent with snowfall or surface deposition, which can form ice with very low dust content. In contrast, pore-filling ice formed by vapor diffusion typically incorporates a higher fraction of dust, since vapor condenses within a porous matrix rather than forming a clean, layered deposit. Because exposed mid-latitude subsurface ice is almost pure (less than 2 percent dust \citep{Dundas2018,Khuller2021,Khuller2021b}), surface deposition followed by sublimation is the more likely emplacement mechanism in these regions. The lag layer acts as both a diffusion barrier and a thermal insulator, both shielding and mitigating subsurface ice sublimation \citep{schorghofer2005,Hudson2007}.

The distribution of mid-latitude subsurface ice has been observed using several methods, including the Odyssey Neutron Detectors \citep{Boynton2002}, the Trace Gas Orbiter Fine-Resolution Epithermal Neutron Detector (FREND) \citep{Malakov2022}, fresh craters \citep{Byrne2009b,Dundas2021,dundas2023}, radar returns \citep{Morgan2021}, surface geomorphology \citep{kreslavsky2002,Morgan2021}, and seasonal surface temperature variations \citep{bandfield2008, Piqueux2019}. Shallow subsurface ice (less than 1 m deep) detected at latitudes below 50\deg{} contrasts with equilibrium models, which predict that under present-day conditions stable ice should exist only poleward of roughly 60\deg{}\citep{schorghofer2005}. A similar problem occurs on Earth in the Dry Valleys of Antarctica, a terrestrial analog of Mars, where there is a discrepancy between the observed depth of the ice table and the predicted equilibrium depth \citep{fisher2016}. If both the equilibrium models and observations are correct, the only option is for the shallow mid-latitude subsurface ice to be out-of-equilibrium and slowly retreating. 

Previous subsurface ice time-marching models often predicted burial depths deeper than observed, primarily because they assumed much earlier emplacement of the ice or a higher dust content \citep{Schorghofer2012,Bramson2017}. These models laid the groundwork for mid-latitude ice sheet evolution using time-averaged solutions for the appropriate physics of subsurface ice sublimation. However, climate responses to orbital forcing are complex, and these models do not incorporate a detailed treatment of atmospheric dynamics, such as diurnal and seasonal cycles of atmospheric water vapor and subsurface–atmosphere interactions, and how these may vary throughout Mars’ past. These prescribed humidity fields tend to underestimate moisture at high obliquity, as the radiative effects of clouds were not taken into account. To test whether the mid-latitude subsurface ice may be out of equilibrium and slowly retreating, as well as to explain its observed spatial distribution, we develop a self-consistent model of subsurface ice sublimation (see Methods). Our approach relies on the state-of-the-art Planetary Climate Model (PCM; \citep{forget1999}), which resolves the diurnal and seasonal cycles of temperature, water, CO$_2$, and dust at a 5-minute resolution (For a full description see Methods). The Mars PCM better simulates the subsurface ice evolution compared to the previous thermal models by including key physical processes such as: latent heat release, albedo changes, atmospheric turbulent mixing, the formation of seasonal frost that alters surface and subsurface temperatures, and the feedback of ice loss on atmospheric humidity. Crucially, it also incorporates the radiative effect of water-ice clouds, which strongly enhances atmospheric humidity at high obliquity \citep{madeleine2014}. The main limitation of our model is that it does not simulate pore-ice growth, although this process has been shown to be negligible on obliquity timescales \citep{vos2023}.

Our simulations start with a surface ice mantle $\sim$630~kyr~ago (or 4.18 Myr ago for Figures SI 5,6,7). This assumption is based on our climate simulations using the Mars PCM that predicts surface ice accumulation in the mid-latitudes at obliquity around 35\deg{} \citep{naar2024recent, madeleine2014}, an obliquity value that last occurred $\sim$630~kyr~ago \citep{laskar2004}. We also tested the case for the last very-high obliquity period (40\deg{}), 4.18 Myr ago, as performed in former studies \citep{Schorghofer2012}. The ice is mixed with a trace amount of dust ($<$2\%) as suggested by spectroscopic analysis of the exposed mid-latitude ice \citep{Dundas2018,Khuller2021,Khuller2021b}. 

\begin{figure*}
\centering
\includegraphics[width=\textwidth]{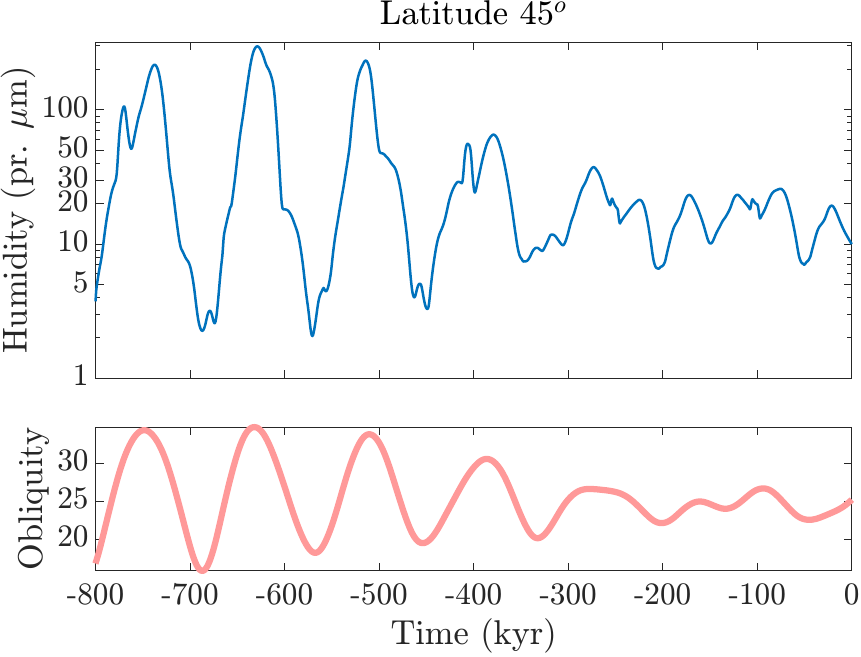}
\caption{Modeled annual zonal mean atmospheric humidity (expressed here by the column water vapor in pr.~$\mu$m) as a function of time before present at latitude 45\deg N, Bottom: Obliquity as a function of time\citep{laskar2004}. Note that the humidity at 45\deg{}N is found to vary by 50-200\% depending on the longitude due to topography and winds.}
    \label{f:hum}
\end{figure*}

As shown in previous studies \citep{schorghofer2005,mellon1993}, the sublimation rate of the subsurface ice is strongly controlled by atmospheric humidity. Hence, to reconstruct the burial history of the ice sheets, we first compute the evolution of the atmospheric humidity over the past few million years at each location on Mars, using an ensemble of 3-D simulations that spans the orbital elements of the last 4 Myr \citep{laskar2004} and assumes a perennial north polar cap as today. As an illustration, Figure~\ref{f:hum} shows the zonal-mean humidity at 45\textdegree N together with the obliquity history. The humidity closely follows the polar insolation at the summer solstice \citep{levrard2007,vos2023}, which is mostly controlled by the obliquity. Higher obliquity values lead to greater atmospheric water content; for example, about 630 kyr ago (when obliquity was $\sim$35\deg{}), the zonal-mean atmospheric water content at latitude 45\deg{} reached 300 $pr. \mu m$, while 570 kyr (when obliquity dropped below 20\deg{}) the water content was just 2.5 $pr. \mu m$.

\begin{figure*}
\centering
\includegraphics[width=\textwidth]{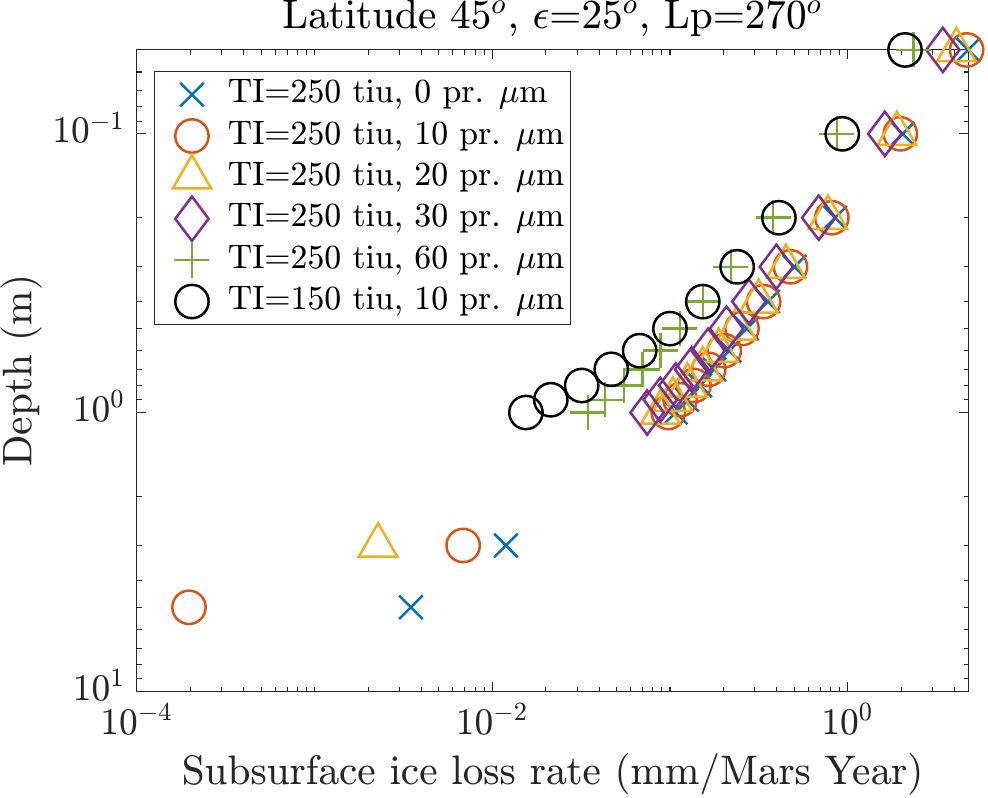}
\caption{Example of subsurface ice loss rate at 45\deg~N modeled by the 1D PCM as a function of depth and atmospheric humidity (expressed here by the column water vapor in pr.~$\mu$m) and the above dry-soil layer thermal inertia, for present-day obliquity and eccentricity, and $L_p$ of 270\deg{}. There is a strong and moderate decrease in flux with depth and atmospheric water content, respectively.}
    \label{f:SSI flux}
\end{figure*}

We then calculate the sublimation rate of the subsurface ice using the 1D version of the PCM  (described in the Methods) for each depth, location, orbital configuration, and humidity. Our nominal choice of model parameters (e.g., wind, regolith diffusion coefficient) is summarized in Table S1, and sensitivity analysis for the atmospheric mixing, wind, and eccentricity are shown in Figures SI 2-4. As an example, Figure~\ref{f:SSI flux} shows the subsurface ice loss rate as a function of depth and humidity at 45\textdegree~N, under present-day obliquity (25.19\deg{}), eccentricity (0.093), and a solar longitude of perihelion $L_p = 270$\textdegree.

Our model predicts a subsurface ice loss rate of 4.5 mm per Mars year at 45\textdegree N and 5 cm depth for present-like conditions. The loss rate decreases with depth, increased atmospheric humidity, and lower surface thermal inertia. 30~centimeters of burial depth is enough to decrease the flux below 1 mm/Mars Year, even for dry atmospheric conditions. The subsurface ice is stable at 1 m depth for atmospheric water contents exceeding 20 pr-$\mu$m—conditions that could occur at slightly higher obliquity or $L_p = 90$ and present-day obliquity\deg{} \citep{vos2023}. The depth dependence arises because (i) diffusion through the regolith buffers the sublimation flux \citep{Hudson2007}, and (ii) the insulating lag reduces the maximum ice temperature, lowering vapor pressure and thus sublimation rates.

\begin{figure}
\centering
\includegraphics[width=\linewidth]{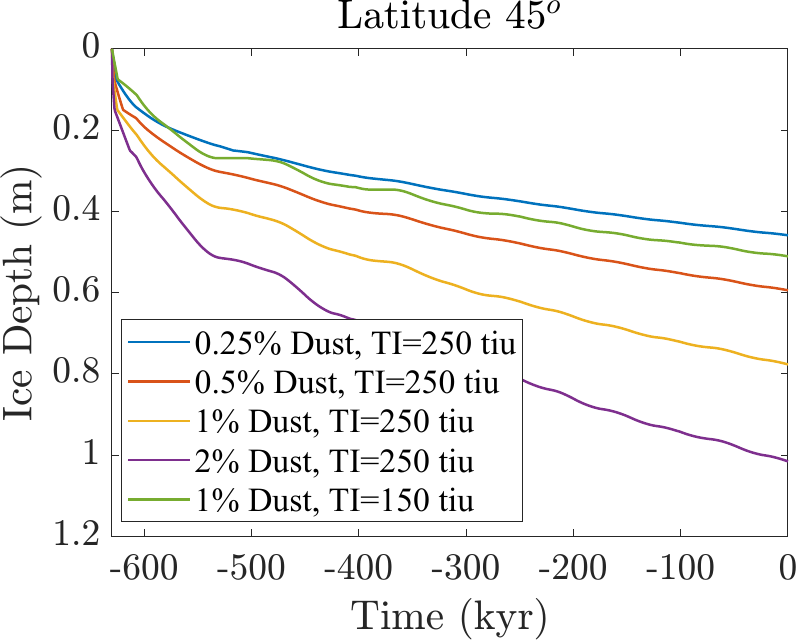}
\caption{Subsurface ice depth modeled evolution as a function of time at latitude 45\deg{} N for emplacement time of 630~kyr ago, no porosity, several initial dust fractions in the ice and lag layers thermal inertia. The ice depth increases as time progresses, with a burial depth of 45-102 centimeters for realistic initial dust fractions in the ice levels and thermal inertia values.}
\label{f:SSI depth}
\end{figure}

To track the evolution of the subsurface ice depth as the orbital configuration evolves (and thus, the surface temperature and humidity), we calculated the subsurface ice loss rate as a function of the orbital elements (an example is shown in Figure SI 1). Given the subsurface ice loss rate as a function of depth, humidity, the orbital configuration history, and assuming the ice sheet deposition time, porosity, and dust content, we integrate the subsurface ice loss rate in time to compute the ice depth as a function of time. Figures~\ref{f:SSI depth}, SI5-7, illustrate the ice evolution for different dust fractions, soil thermal inertias, and porosities, as well as for two initial surface ice deposition times: 630 kyr ago and 4.18 Myr ago, corresponding to the last occurrences of 35\textdegree and 40\textdegree obliquity, respectively. The retreat rate of the subsurface ice decreases as the subsurface ice depth becomes larger because of the two mechanisms previously mentioned. We consider our model to provide an upper limit on the depth to the top of the ice sheet, because in our model the subsurface ice sheet can only retreat; however, in reality, when the ice table is stable, pore-filling ice may grow in the medium between the ice table and the surface \citep{schorghofer2005,Schorghofer2007,Schorghofer2012,Bramson2017}, and later mitigate the ice table loss. Previous work \citep{vos2023} demonstrated that pore ice requires million-year timescales to accumulate in significant amounts; thus, its contribution is expected to be minor. Nevertheless, our calculations represent an upper bound on the ice depth.

The subsurface ice final depth for the assumed parameters shown in Figure 3 is between 0.45 and 1.02 meters. The large difference between 1\% dust with thermal inertias of 250 and 150 tiu shows that the ultimate ice depth is strongly influenced by this parameter (or more precisely by the related thermal conductivity of the lag layer) due to its control of the subsurface maximum temperature.

\begin{figure*}
\centering
\includegraphics[width=\textwidth]{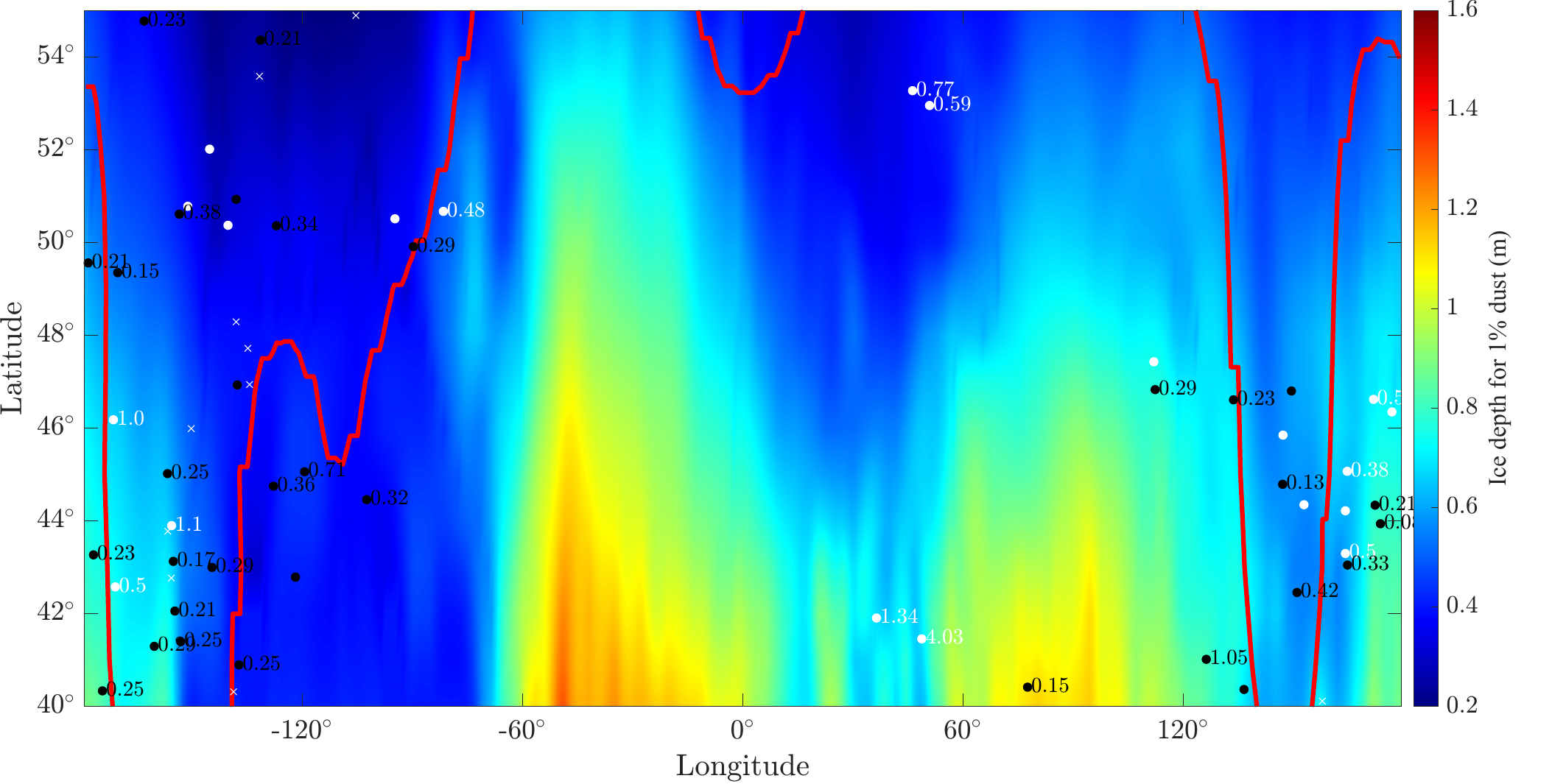}
\caption{Subsurface ice final depth interpolated map for latitudes 40-55 North, assuming ice was emplaced 630 kyr ago for 1\% dust fraction scenario and no lag layer porosity. The red line represents MONS 7.5\% Water-Equivalent Hydrogen contour\citep{PATHARE2018}. The white circles and crosses are locations where ice-exposing craters and maybe ice-exposing craters are found, respectively. The black circles are locations where non-ice-exposing craters are found. The numbers next to the circles indicate the assumed crater depths \citep{Dundas2021}.}
\label{f:SSI depth_lat}
\end{figure*}

We mapped the modeled depth of the ice in the latitudinal belt between 40\deg{} N and 55\deg{} N (with a 1\textdegree steps), taking into account the observed variation in surface thermal inertia\citep{Putzig2005} assumed here to be the thermal inertia of the entire lag layer. Figure~\ref{f:SSI depth_lat} presents this map of the final depth of the subsurface ice assuming that it was placed 630~kyr ago with 1\% dust fraction and no porosity.  Figure~\ref{f:SSI depth_lat} also shows the location of the observed ice-exposing and non-ice-exposing craters \citep{Dundas2021} and the 7.5\% Water-Equivalent Hydrogen contour measured by the MONS Neutron spectrometer, which is a proxy for shallow subsurface ice \citep{PATHARE2018}.  We find that the modeled ice depth for these assumptions is between 20 and 150 centimeters, with substantial longitudinal variations that agree well with the Mars Odyssey Neutron Spectrometer (MONS) observations \citep{PATHARE2018}. Overall, ice depth becomes shallower at higher latitudes and increases with thermal inertia, reflecting the combined influence of surface temperature gradients and regolith properties on ice stability. The model predictions and observations are in good agreement: ice-exposing crater depths exceed the modeled ice depth, indicating penetration into the ice sheet, whereas non-ice-exposing crater depths are shallower than the modeled values, consistent with the absence of excavated ice. In some locations, HiRISE observations found ice-exposing craters at depths larger than 1~m \citep{Dundas2021} where MONS did not measure high Water-Equivalent Hydrogen, for example, around latitude 42$\deg{}$ north and longitude 50$\deg{}$. Our model results suggest that the ice depth in these locations is $\sim$~70 centimeters, which is outside the MONS range ($\sim$~50~cm) but inside the ice-exposing craters range (as can be inferred from the craters' depths).
Figure SI8 presents a similar map to Figure 4, assuming that it was placed 4.18~Myr ago with 1\% dust fraction and no porosity.

In this paper, we show that the subsurface ice observed at mid-latitudes below 55\textdegree is likely out of equilibrium, representing slowly retreating surface-deposited ice. Our calculations indicate that if emplacement occurred ~630 kyr ago—when obliquity last reached 35\textdegree and favored mid-latitude accumulation—the predicted depths are consistent with both the observed values and their spatial variability \citep{Dundas2021,PATHARE2018}. In particular, craters that expose ice are deeper than the modeled ice table, whereas non-exposing craters are shallower and do not reach it. Because the ice loss rate decreases sharply with depth (Figures \ref{f:SSI flux}, \ref{f:SSI depth}, SI1,5-7), earlier emplacement ages cannot be excluded. However, assuming deposition 4.18 Myr ago results in weaker agreement with observations (Figure SI8). Taken together, these results suggest that mid-latitude surface ice was last emplaced more recently than 4 Myr ago, in good agreement with estimates from geomorphological analyses \citep{Head2003,Kostama2006,Schon2012}.

Because the sublimation loss rate drops sharply with depth, even a more porous regolith lag layer has only a modest effect on the resulting depth of the subsurface ice. (compare Figure 3 with Figure SI5). In this case, the lag layer develops more rapidly, resulting in greater lag layer thickness, which slows the ice retreat. To form the observed dust-lag deposits, the thickness of sublimated ice (assuming no surface dust removal or addition, and no ice porosity) is given by $Z_{ice}\frac{1-\phi}{f_{dust}}$, where $Z_{ice}$ amount of ice loss, $\phi$ is the regolith lag layer porosity, and $f_{dust}$ is the dust fraction in the ice (see Methods). For example, sublimating a 5-meter ice layer would produce a 10-centimeter-thick lag deposit (assuming a lag porosity of 0\% and a 2\% dust content in the lost ice).  
Even with a conservative deposition rate of 1 mm per year, which is lower than estimates from previous studies \citep{madeleine2014,vos2022,naar2024recent}, a 5-meter ice sheet would form in about 5000 years, well within a reasonable timescale. Taken together, these results indicate that the mid-latitude subsurface ice most likely represents remnants of a thick, orbitally emplaced ice sheet that has been retreating since its last major accumulation. This also highlights that the mid-latitude subsurface ice preserves a record of past surface ice sheets, providing key constraints on the timing, magnitude, and retreat of the climate-driven water cycle.

\section*{Acknowledgements} 
This work is funded by the European Research Council (ERC) under the European Union’s Horizon 2020 research and innovation program (grant \#408 835275, project ``Mars Through Time''). Mars PCM simulations were done thanks to the High-Performance Computing (HPC) resources of Centre Informatique National de l'Enseignement Supérieur (CINES) under the allocation n\textdegree A0100110391 made by Grand Equipement National de Calcul Intensif (GENCI); and thanks to the IPSL Data and Computing Center ESPRI, which is supported by CNRS, Sorbonne Université, CNES and Ecole Polytechnique.
The authors would like to thank Ali Bramson and the two anonymous reviewers for their helpful comments and suggestions, which improved this manuscript.

\vfill\null

\bibliographystyle{naturemag}
\bibliography{Mars}

\end{document}